\documentclass[10pt,conference]{IEEEtran} 

\IEEEoverridecommandlockouts

\usepackage{cite}
\usepackage{amsmath,amssymb,amsfonts}
\usepackage{algorithmic}
\usepackage{graphicx}
\usepackage{textcomp}
\usepackage{subcaption}
\usepackage{xcolor}

\usepackage{booktabs}

\usepackage{todonotes}

\def\BibTeX{{\rm B\kern-.05em{\sc i\kern-.025em b}\kern-.08em
    T\kern-.1667em\lower.7ex\hbox{E}\kern-.125emX}}
\begin{document}

\title{Quantum Circuit Caches and Compressors\\for Low Latency, High Throughput Computing}

\author{
    \IEEEauthorblockN{
        Ioana Moflic\IEEEauthorrefmark{1},
        Alan Robertson\IEEEauthorrefmark{2}, 
        Simon J. Devitt\IEEEauthorrefmark{2}\IEEEauthorrefmark{1} and
        Alexandru Paler\IEEEauthorrefmark{1},\IEEEauthorrefmark{3}
    }
    \IEEEauthorblockA{
        \IEEEauthorrefmark{1}\textit{Aalto University, Helsinki, Finland},\\
        \IEEEauthorrefmark{2}\textit{University of Technology Sydney, Australia},\\
    }
}

\maketitle

\begin{abstract}
Utility-scale quantum programs contain operations on the order of $>10^{15}$ which must be prepared and piped from a classical co-processor to the control unit of the quantum device. The latency of this process significantly increases with the size of the program: existing high-level classical representations of quantum programs are typically memory intensive and do not na\"ively efficiently scale to the degree required to execute utility-scale programs in real-time. To combat this limitation, we propose the utilization of high-level quantum circuit caches and compressors. The first save on the time associated with repetitive tasks and sub-circuits, and the latter are useful for representing the programs/circuits in memory-efficient formats. We present numerical evidence that caches and compressors can offer five orders of magnitude lower latencies during the automatic transpilation of extremely large quantum circuits.
\end{abstract}


\section{Introduction}

Quantum-HPC (QHPC) infrastructures are built by interconnecting quantum computers with classical high-performance computers (HPC)~\cite{suchara2018hybrid, humble2021quantum}. QHPCs are expected to use the power of HPCs to control the quantum computation and at the same time, the classical parts of the quantum computations are offloaded to the HPC.

Both latency and throughput of the HPC$\rightarrow$quantum connection play a crucial role in the success of executing a full quantum computation. This connection is used for communicating operations abstracted in the form of quantum circuits (e.g. OpenQASM 3.0~\cite{cross2022openqasm}), which include quantum gates, quantum measurements and classical signals (e.g. classically controlled quantum gates). The connection can also be used by the control electronics~\cite{battistel2023real, niu2019universal} to send simple classical control sequences to the quantum system. The quantum$\rightarrow$HPC connection communicates the classical bits of the mid-circuit measurements. These bits are used for partial reconstructions of the states~\cite{aaronson2018shadow}, quantum error-mitigation~\cite{cai2023quantum} or -correction.

The execution of computations on a QHPC is controlled by a quantum operating system~\cite{paler2024architecting, giortamis2024qos} (QOS), which will operate using the execution model of a classical \emph{just-in-time} (JIT) compiler. The QOS takes a high-level quantum circuit description of the computation, analyzes it in order to allocate the necessary computational resources, automatically partitions the whole computation to sub-circuits, and processes the partitions before sending these for execution. The quantum computer is not necessarily monolithic and can be formed from multiple networked quantum processing units~\cite{caleffi2024distributed, van2016path, saadatmand2024fault} running independent partitions.

Partitions are a practical approach, since most quantum computations have repeated sub-circuits/partitions. For example, the query of a QROM/QRAM is a sub-circuit where only a few parameters are changed, or the majority gate (MAJ) in carry-ripple adders~\cite{cuccaro} appears in a very regular pattern.

Processing a partition is a time-consuming task and involves transpiling the partition to the native gate set of the quantum computer~\cite{baker2020time, brandhofer2023optimal}, further optimizing it to the constraints of the quantum computer, or encoding it for transmission. The QOS manages the execution of the partitions and is, in general, responsible for the successful execution of the entire quantum computation. To this end, the QOS will also monitor and manage communication between interconnects.

\begin{figure}[!t]
    \centering
    \includegraphics[width=0.8\columnwidth]{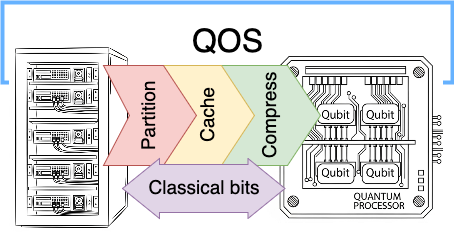}
    \caption{A quantum operating system (QOS) is controlling the QHPC. Classical bits (e.g. from mid-circuit measurements and simple control signals) are sent both ways between the HPC and quantum computer. Introducing quantum circuit caches and compressors on the interconnect between the HPC and the quantum systems improves the latency and throughput of executing large quantum computations. Note that depending on the representation of the quantum circuits, one can cache before compressing or vice versa.}
    \label{fig:arch}
    \vspace{-1em}
\end{figure}

\section{Low Latency, High Throughput Execution of Quantum Computations}

The entire computation, and for that matter the partitions too, are executed in a permanent feedback loop, where the HPC system is \emph{streaming} the processed partitions/sub-circuits to the quantum computer, while simultaneously receiving streams of classical bits from the quantum computer. In practice, the QOS operates a pipeline (Fig.~\ref{fig:arch}). 

Practical quantum circuits, once decomposed into quantum error-correction primitives, include more than $10^{15}$ gates, e.g.~\cite{caesura2025faster}. Therefore, in order to execute the entire computation, it is necessary to compile it just-in-time while making sure that it does not degrade the performance of the execution. In this context, there is a need for high-performing tools that can compute, process, and stream partitions to the quantum computer fast. 

The HPC$\rightarrow$quantum connection might experience \emph{high latency} if the quantum computer waits for the HPC to compute the next partition to send. The first challenge is to compute partitions very fast. The second challenge is to minimize the processing time of a partition. The compiler must ensure that, if a similar partition has already been processed, it reuses the cached result by sending it directly to the quantum computer, thereby saving processing time.

The HPC$\rightarrow$quantum connection might experience \emph{low throughput} if inefficient data formats are used to communicate the partitions. For example, the OpenQASM string representation of a circuit may be too verbose such that for a fixed connection bandwidth, the rate of communicated quantum gates/second is low.

To address the previous challenges, we build on top of two tools: Pandora and Cabaliser. Pandora~\cite{moflic2024scalable} is integrated with Google Qualtran and pyLIQTR and can transpile utility-scale circuits such as quantum chemistry or Shor's algorithm to Clifford+T. Pandora can also process quantum circuits and can operate as a cache for these. Pandora is built using a relational database system, such that it is natively multi-threaded and most tasks can easily be parallelized due to data-consistency guarantees that relational databases give in concurrent scenarios.

Cabaliser can compile from Clifford + T to a graph state representation of the program~\cite{Krishnan_Vijayan_2024, Aaronson_2004}. 
Graph-state representations of quantum circuits consist of a reduction of the Clifford components of the circuits to a graph and a set of local Clifford gates\cite{schlingemann2001stabilizercodesrealizedgraph} and the non-Clifford components to a sequence of teleported $Rz(\theta, \varepsilon)$ measurements. 
Cabaliser can encode graph state circuits in very efficient serialisable bit-packed structures.

In the following, we detail how low latency (Sections~\ref{sec:partition} and~\ref{sec:cache}) and high throughput (Sections~\ref{sec:compress}) can be achieved using Pandora and Cabaliser.

\subsection{Partitioning Quantum Circuits}
\label{sec:partition}

The streamed partitions have to be compatible with constraints imposed by the quantum computer, such as a maximum T-count, maximum gate depth or number of qubits.

Pandora stores circuits internally as directed acyclic graphs (DAG) which are processed and optimized by custom rewrite rules via pattern matching. The rewrites can be performed both sequentially (i.e. iterating in topological order over the DAG) and randomly (i.e. accessing gates in random locations of the DAG). We implemented a partitioning algorithm on top of Pandora. Our algorithm leverages union-find data structures and works as follows:
\begin{enumerate}
    \item we begin by generating the topologically-ordered edge list of $(V_i, V_{i+1})$ and cache the list into Pandora, where each $V_k$ is a node of the DAG. Each element of the list is identified with a directed edge from node $V_i$ to node $V_{i+1}$. This step is performed only once.
    
    \item we continue by applying the \texttt{union} and \texttt{find} operations and grow node partitions as long as no bounds are exceeded (e.g. the T-count of the partition is lower or equal than the maximum allowed).
    
    \item we return the generated partitions.
\end{enumerate}

The union-find algorithm has a space complexity of $\mathcal{O}(n)$, where $n$ is the number of nodes in the DAG and a time complexity of $\mathcal{O}(\alpha(n))$, which is practically constant even for very large inputs -- $\alpha$ grows extremely slowly. 

Due to space being a possible limitation, we never perform the in-memory partitioning algorithm on the whole DAG, but rather on time windows of the DAG. We extract batches from the cached edge list and only perform partitioning on one batch at a time in order to reduce the memory footprint.

\begin{figure}[!t]
    \centering
    \includegraphics[width=0.9\columnwidth]{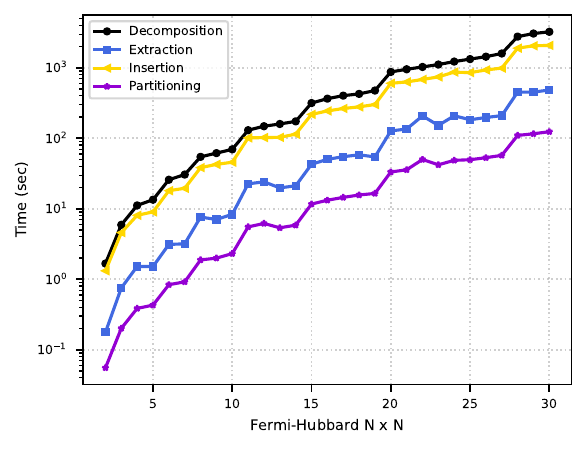}
    \caption{JIT compilation for Fermi-Hubbard circuits. Durations for the Pandora pipeline in order of execution: decomposition time in \texttt{pyLIQTR}, insertion, partitioning and extraction times.}
    \label{fig:pandora_steps}
    \vspace{-1em}
\end{figure}

\begin{figure}[htbp]
    \centering
    \includegraphics[width=0.9\columnwidth]{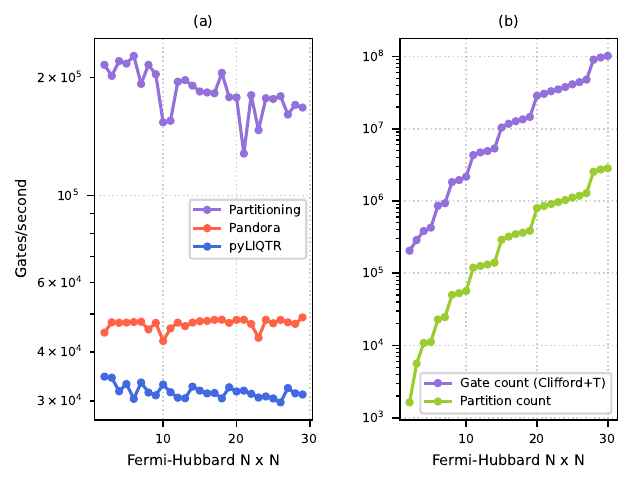}
    \caption{Gate processing speeds and number of partitions: (a) The number of gates processed per second at different stages of the pipeline - Once a circuit is decomposed and stored, the partitioning algorithm achieves very high speeds; (b) The gate processing speed is decreasing slightly with increasing Fermi-Hubbard circuit sizes, while the number of partitions increases monotonically. This is to show that the bandwidth of the HPC$\rightarrow$quantum connection is not saturated and even larger circuits, and more partitions could be streamed while maintaining high gate streaming speeds.}
    \label{fig:widgetstats}
    \vspace{-1em}
\end{figure}

\subsection{Caching Quantum Circuits}
\label{sec:cache}

Caches can help improve the performance and latency of an application by storing frequently accessed data in fast memory storage.

In Pandora, circuit transpilation and insertion, optimisation and extraction can all be performed using a variable number of parallel processes to increase processing speed. Due to little or no communication that the processes have to perform, the speed-up from paralellizing such tasks is close to linear. 

Low-latency transpilations can be achieved by:
\begin{enumerate}
    \item taking the repetitive structure of the circuit into account and decomposing only sub-circuits which are not already cached into Pandora (e.g if a circuit contains multiple adders, we only transpile a specific n-bit adder once -- Section~\ref{sec:adder}),
    
    \item using multi-threaded transpilation of the circuit's DAG. The speed-up is obtained by assigning sub-DAGs to separate transpilation processes.
\end{enumerate}

Transpilation is performed in a streamlined fashion, in order to keep the memory requirements low. Each transpilation process decomposes a fixed-size batch of gates and inserts it as a single entity. After insertion, the process frees the memory associated with the batch and proceeds with the next. Ideally, the number of inserts into the database is kept to a minimum because they are computationally expensive. Depending on the available hardware, a trade-off between the number of processes and the batch size has to be found in order to maximize performance.

\subsection{Compressing Quantum Circuits}
\label{sec:compress}

Given the large scale of the input program, we require that operations are streamed in logical chunks (partitions) that may be processed in sequence.
Allocating and processing blocks of instructions of a bounded logical size reduces variability between computations by imposing a maximum logical footprint for each chunk. 
In practical terms, this allows for the use of slot allocators rather than relying on memory arenas and calls to the system allocator.

The graph state of a partition representation bounds the Clifford components of an $n$ qubit input circuit with $k$ non-Clifford gates to the preparation of a graph state - describable by at most $(n + k)^2$ pairwise entries along with $(n + k)$ local Clifford operations. 
For some circuits $k$ is large rendering this scaling is initially infeasible.
However, by splitting the input circuit and stitching the compressed graph states we can enforce a constant bound on the number of elements of any given graph.  
Teleporting input qubits to perform graph stitching increases the size of the graph by $n$ elements.

The implementation of non-Clifford elements of the circuit may require representations up to algebraic precision.
To maintain these graph tokens as precision-agnostic constructs we do not attempt to serialise arbitrary precision floating point operations - instead we construct a tagged cache table of known decomposition sequences and serialise the appropriate key as an integer. This is a caching technique in addition to the one from Section~\ref{sec:cache}. The decomposition of each non-Clifford operation reduces to a sequence of $T_{\sigma}$ operations that may be cached and serialised independently.

This reduces the total compilation target object to $(2n + k)^2$ graph edge pairs, up to $n + k$ local Cliffords, and up to $n + k$ non-Clifford sequence cache keys.

By selecting a fixed quantum co-processor operation cache size, we may then bound the graph state chunk size as defined by the number of logical qubits and the number of non-Clifford gates.  
The Union-Find partitioning implemented may then be constrained based on these parameters to only emit partitions that satisfy these bounds.

The internal tableau structures of Cabaliser are a set of dense bit-matrices, which are manipulated to compile graph states using vector (AVX2, BMI2) operations\cite{Aaronson_2004}.
The graph state is then emitted as a collection of bit-packed tables as described above. This bit-packed graph state compilation acts as a compression function that reduces the input quantum circuit to a linear sequence of graph tokens of constant bounded size that correspond to memory-local operations. As the stitched graph states are agnostic to the logical qubits that they act over, these graph tokens may be also be cached and re-emitted.

\section{Results}

\subsection{Fermi-Hubbard}

We show the advantage of caching in Fig.~\ref{fig:pandora_steps}, which illustrates the performance of compiling Fermi-Hubbard circuits of varying sizes in the Pandora pipeline. \emph{Insertion} is the process of inserting a circuit generated by pyLIQTR, or any other software, into Pandora. The insertion time is dominated by disk read and write times. \emph{Extraction} is the process of exporting circuits from Pandora. There is a distinction between extraction and partitioning. The first refers to the process of assembling the Pandora information into a standardized format such as OpenQASM, whereas the latter is the process of finding architecture-aware sub-circuits.

We record the durations of each pipeline stage: the transpilation from \texttt{pyLIQTR} (black - circle) and the insertion (yellow - triangle) are the most time-consuming stages. Partitioning (blue - square) and extraction (purple - star) are the most efficient stages. In practice, partitioning and extraction are repeated multiple times, achieving low latency HPC$\rightarrow$quantum connections (cf. Fig.~\ref{fig:widgetstats} and Section~\ref{sec:adder}).

\subsection{Arithmetic}
\label{sec:adder}

Na\"ively, the cost of assembling addition circuits scales linearly in the number of qubits.   
Using Pandora as a cache, we compose a strided MAJ and UMA gate~\cite{cuccaro} to a constant stride.
This forms a simple example of a circuit with a native partitioning that admits a non-trivial degree of partition reuse.

We bound this stride by the graph state window size $\alpha$.
The action of any $\alpha k$-qubit adder may then be expressed by $k$ repetitions of the strided MAJ and UMA objects.  
More general $\beta + \alpha k$-bit adders can be constructed by caching a second set of UMA and MAJ objects with a stride $\beta$.  

The circuit for an arbitrary depth adder can then be expressed in terms of four graph states with constant memory up to the permutation of input qubits for each graph instance.

The performance of the construction of these circuits can be seen in Table \ref{tab:adder}. As expected, the runtime of the strided decomposition only depends on the stride, while the circuit construction cost and decomposition for the full adder circuits scales with the number of bits.

For both strided and full decompositions, the Pandora insertion and decomposition time was typically 5\% of the reported runtime, with the remainder of runtime occurring during the initial circuit construction in Python. This strategy enables the fast computation of utility scale quantum algorithms from circuit partitions and supports a JIT execution model.

\begin{table}
\setlength{\tabcolsep}{4pt}
\begin{center}
\begin{tabular}{rrrcrc} 
\toprule
Adder & Stride & Full Dec. & Strided Dec. & Full Gr. & Strided Gr. \\
(Bits) & (Bits) &  & &  & \\  \midrule
128  & 64 & $1.79 \pm 0.01$ & $0.69 \pm 0.02$ & $0.56 \pm 0.1$ & $0.007$\\ 
256  & 64 & $5.09 \pm 0.03$ & $0.70 \pm 0.01$ & $2.2 \pm 0.2$ & $0.008$\\ 
512  & 64 & $18.3 \pm 0.70 $ & $0.70 \pm 0.01$ & $13.9 \pm 1.3$ & 0.008\\  
1024 & 64 & $74   \pm 2.50 $ & $0.70 \pm 0.01$ & $92 \pm 3.8$ & 0.007\\ 
2048 & 64 & $271  \pm 8.00 $   & $0.71 \pm 0.01$ & $527 \pm 24.9$ & 0.007\\ 
128  & 128  & $1.79 \pm 0.01$ & $1.58 \pm 0.04$ & $0.6 \pm 0.1$ & 0.030\\ 
256  & 128  & $5.09 \pm 0.03$ & $1.55 \pm 0.01$ & $2.1 \pm 0.2$& 0.031\\ 
512  & 128  & $18.3 \pm 0.70 $ & $1.56 \pm 0.03$ & $13.7 \pm 1.6$ & 0.029\\  
1024 & 128  & $71   \pm 0.60 $ & $1.54 \pm 0.01$ & $89.7 \pm 1.9$& 0.030\\ 
2048 & 128  & $284  \pm 8.20 $ & $1.54 \pm 0.01$ & $527 \pm 24.9$& 0.030

\end{tabular}
\end{center}
\caption{Average times in seconds for single threaded circuit decomposition with Pandora working as a cache and strided graph state constructions against full circuit decompositions. Choosing a stride of 64 bits allows for the decomposition and construction of all graph states required to implement a 2048 bit adder in less than 1 second. Results were obtained on an Intel i7-1185G7 chip. For 2048-bit adders and strides of 64, the speedup is five orders of magnitude ($\frac{527}{0.007}$.)}
\label{tab:adder}
\vspace{-1em}
\end{table}

\section{Conclusion}
We introduced and demonstrated the efficacy of high-level quantum circuit caches and compressors as crucial components for achieving low-latency, high-throughput quantum computing within Quantum-HPC infrastructures.

Our system integrates Pandora for caching and partitioning and Cabaliser for efficient graph-state compression. We presented numerical evidence that these techniques can yield a five-orders-of-magnitude reduction in latencies during the automatic transpilation of extremely large quantum circuits.

\section*{Acknowledgements}
This research was developed in part with funding from the Defense Advanced Research Projects Agency [under the Quantum Benchmarking (QB) program under award no. HR00112230006 and HR001121S0026 contracts]. The views, opinions and/or findings expressed are those of the author(s) and should not be interpreted as representing the official views or policies of the Department of Defense or the U.S. Government.

\bibliographystyle{unsrt}
\bibliography{__main}

\end{document}